\documentclass[pra,reprint,twocolumn,superscriptaddress,showpacs,floatfix]{revtex4-1}
\usepackage{amsmath}
\usepackage{mathrsfs}
\usepackage{boondox-cal}
\usepackage{txfonts}
\usepackage{amssymb}
\usepackage{graphicx}
\usepackage{hyperref}
\usepackage{ulem}
\usepackage{xcolor} 
\usepackage{color}
\usepackage{overpic}
\usepackage{psfrag}
\usepackage{tabularx}
\usepackage{multirow}
\usepackage{array}
\usepackage{placeins}
\newcommand{\PreserveBackslash}[1]{\let\temp=\\#1\let\\=\temp}

\usepackage{bm}
\usepackage{dsfont}
\usepackage{mathtools}

\makeatletter

\begin{document}

\title{Non-invertible symmetries and boundary conditions for the transverse-field Ising model}

\author{Huan-Qiang Zhou}
\affiliation{Centre for Modern Physics, Chongqing University, Chongqing 400044, The People's Republic of China}

\author{Qian-Qian Shi}
\affiliation{Centre for Modern Physics, Chongqing University, Chongqing 400044, The People's Republic of China}

\begin{abstract}
Non-invertible Kramers-Wannier (KW)  duality symmetries are constructed for the transverse-field Ising model (TFIM) at the self-dual point under various boundary conditions (BCs), as long as the resultant Hamiltonian commutes with the ${\rm Z}_2$ symmetry operator.  This is achieved by introducing extra degrees of freedom into the Hilbert space,  in order to turn a non-translation-invariant Hamiltonian in the original Hilbert space into a translation-invariant Hamiltonian in the augmented Hilbert space. One may lift the trivial identity operator, the ${\rm Z}_2$ symmetry operator and the non-invertible KW  duality symmetry operator to their counterparts in the augmented Hilbert space, valid for each of four types of toroidal BCs. As it turns out, they yield a lattice version of fusion rules, which bears a resemblance to the Tambara-Yamagami ${\rm Z}_2$ fusion category. Our construction is thus consistent with the basic physical requirement that all possible BCs should yield a converging result in the thermodynamic limit. In particular, the lattice versions of fusion rules, constructed by Seiberg, Seifnashri and Shao [SciPost Phys.  \textbf{16}, 154 (2024)], are reproduced for periodic and anti-periodic BCs, but a discrepancy is revealed for duality-twisted BCs.
\end{abstract}
\maketitle

There has been revived interest in the well-known Kramers-Wannier (KW) dualities~\cite{kramers} for the transverse-field Ising model (TFIM)~\cite{ising} and its various generalizations, including the $q$-state  quantum Potts (QP) model~\cite{baxterbook,qpotts1},  largely due to the fact that duality is not a symmetry operation in the conventional sense~\cite{seiberg0}. This has led to conceptual developments regarding non-invertible symmetries~\cite{seiberg,shao} in the context of fusion category~\cite{ty} (also cf.~Refs.~\cite{category1,category3,category4,category5}). Namely, the non-invertible KW duality symmetry does not possess an inverse (also cf.~Ref.~\cite{oshikawa} for an alternative treatment).
This is well beyond the conventional realm of symmetry -- a notion well described by an associative group multiplication operation that
obeys a group composition rule such that there is always an inverse for each symmetry group element, including an identity.

Actually,  non-invertible KW duality symmetries are closely related with topological defects on a lattice~\cite{fendley}. As stressed in Refs.~\cite{seiberg,shao}, the topological property of a defect means that the location of the defect is arbitrary and can be varied by conjugating the defect Hamiltonian with local unitary operators. This dichotomy between topological and non-topological defects results in a subtle difference that a lattice version of fusion rules {\it only} arises from a topological defect~\cite{seiberg,shao}. According to Ref.~\cite{seifnashri}, this  reflects a one-to-one correspondence between topological defects and symmetry operators. As a consequence, for the TFIM under toroidal boundary conditions (BCs), including periodic BCs and anti-periodic BCs, non-invertible KW duality symmetries
mix with a variant of the translation symmetry operator and are thus not a fusion category. However, it flows to the Tambara-Yamagami ${\rm Z}_2$ fusion category in the thermodynamic limit, given that the translation symmetry operator becomes trivial. In contrast, no non-invertible KW duality symmetry accompanies the TFIM under non-toroidal BCs.

A natural question arises as to whether or not it is possible to construct a non-invertible KW duality symmetry for the TFIM with free ends -- one of  non-toroidal BCs that are not translation-invariant, in contrast to toroidal BCs, including periodic BCs, anti-periodic BCs and duality-twisted BCs. If the answer were not affirmative, then we would be confronted with the basic physical requirement that  all possible BCs should yield a converging result in the thermodynamic limit. More precisely, the absence of a non-invertible KW duality symmetry for the TFIM with free ends would lead to a serious logical problem, in the sense that the two limiting operations do not commute with each other: one is the limit to reach the TFIM with free ends in a parameter space in which various BCs are parameterized, and the other is the thermodynamic limit.
According to Berry~\cite{berry}, this implies that the thermodynamic limit is singular with respect to a limiting operation in this parameter space. Indeed, for the TFIM under toroidal BCs, the fusion rules of the underlying CFT follow from a lattice analogue, but this scenario fails under non-toroidal BCs, when the thermodynamic limit is reached. The same argument even applies to the situation that a KW duality symmetry is non-invertible for periodic and anti-periodic BCs, but invertible for duality-twisted BCs.

It is tempting to address this intriguing question for the TFIM under various BCs.  A key idea is that, for the TFIM under any BCs, it is possible to  introduce extra degrees of freedom into the original Hilbert space to lift the trivial identity operator, the ${\rm Z}_2$ symmetry operator and the non-invertible KW  duality symmetry operator to their counterparts in the augmented Hilbert space for each of four types of toroidal BCs. As a result, a non-translation-invariant Hamiltonian in the original Hilbert space is turned
into a translation-invariant Hamiltonian in the augmented Hilbert space, as long as the resultant Hamiltonian commutes with the ${\rm Z}_2$ symmetry operator. This trick makes it possible to construct non-invertible KW duality symmetries for the TFIM at the self-dual point under non-toroidal BCs. It follows that the trivial identity operator, the ${\rm Z}_2$ symmetry operator and the non-invertible KW  duality symmetry yield a lattice version of fusion rules in the augmented Hilbert space, which bears a resemblance to the Tambara-Yamagami ${\rm Z}_2$ fusion category~\cite{ty}. This {\it not only} resolves an apparent contradiction with the basic physical requirement that all possible BCs should yield a converging result in the thermodynamic limit, {\it but also} clarifies the essential difference between topological and non-topological defects  in the context of non-invertible KW duality symmetries. 

Our construction shows that the lattice versions of fusion rules, constructed by Seiberg, Seifnashri and Shao in Ref.~\cite{seiberg}, for toroidal BCs, are peculiar, in the sense that they are strongly tied with the translation invariance in the original Hilbert space, thus leading to the involvement of  a variant of the translation operator in a lattice analogue of fusion rules for  each of four types of toroidal BCs.
In particular, we are capable of reproducing their results for both periodic and anti-periodic BCs, but there exists a discrepancy for duality-twisted BCs.

{\it The TFIM under various BCs.~--} 
The TFIM under investigation is described by the Hamiltonian
\begin{equation}
H_{\rm}(\alpha,\beta,\gamma;\lambda)=-\sum_{j=1}^{L-1} \sigma_j^x \sigma_{j+1}^x- \lambda \sum_{j=1}^{L-1} \sigma_{j}^z-B_{L,1}(\alpha,\beta,\gamma;\lambda),  \label{Ising} 
\end{equation}
where $\sigma_i^x$ and $\sigma_i^z$ denote the spin-$1/2$ Pauli matrices at the lattice site labeled by $j$ ($j=1,2,\ldots,L$), with $L$ being the size, and $B_{L,1}(\alpha,\beta,\gamma;\lambda)$ denotes a boundary term describing various BCs that only involve spin degrees of freedom located at the two lattice sites labeled by $L$ and 1. Note that the model has been extensively investigated~\cite{qPotts,levy}.
Here we are only interested in the self-dual point $\lambda=1$, meaning that the bulk is at criticality. From now on we denote $H_{\rm}(\alpha,\beta,\gamma;\lambda=1)$ and $B_{L,1}(\alpha,\beta,\gamma;\lambda=1)$ as $H_{\rm}(\alpha,\beta,\gamma)$ and $B_{L,1}(\alpha,\beta,\gamma)$ for brevity. We remark that the Hilbert space is isomorphic to $\mathbb{C}^2_1 \otimes \mathbb{C}^2_2 \otimes \ldots \otimes \mathbb{C}^2_L$, where $\mathbb{C}^2_j$ ($j=1,2,\dots,L$) denote the two-dimensional complex vector space with an inner product at the lattice site $j$.

Our discussion below applies to any choice of the boundary term $B_{L,1}(\alpha,\beta,\gamma)$, as long as the Hamiltonian (\ref{Ising}) possesses the ${\rm Z}_2$ symmetry generated by $\eta=\prod_{j=1}^L\sigma_{j}^z$. Note that $\eta^2=I$, where $I$ denotes the identity operator in the Hilbert space. However,
we shall focus on a specific choice that only involves spin degrees of freedom located at the two lattice sites labeled by $L$ and 1, namely the boundary term $B_{L,1}(\alpha,\beta,\gamma)$ takes the form: $B_{L,1}(\alpha,\beta,\gamma) = \alpha\sigma_L^x \sigma_{1}^x+\beta \sigma_L^y \sigma_{1}^x+\gamma\sigma_L^z$.	At $\alpha=0$, $\beta=0$ and $\gamma=1$,  the Hamiltonian becomes the critical TFIM with free ends. Further, it corresponds to the critical TFIM under periodic BCs at $\alpha=1$, $\beta=0$ and $\gamma=1$, and to the critical TFIM under anti-periodic BCs at  $\alpha=-1$, $\beta=0$ and $\gamma=1$, respectively.  Meanwhile, at $\alpha=0$, $\beta=0$ and $\gamma=0$,  the Hamiltonian becomes the critical TFIM with free ends, with $-\sigma_L^z$ being absent.  Further, it corresponds to the critical TFIM under duality-twisted BCs at $\alpha=0$,  $\beta=1$ and $\gamma=0$, and to the  critical TFIM under anti-duality-twisted BCs at  $\alpha=0$, $\beta=-1$ and $\gamma=0$, respectively. 

Note that the above six specific BCs are special, because the resultant Hamiltonian in each case possesses an extra conserved quantity. Among them, the Hamiltonian $H_{\rm}(\alpha,\beta,\gamma)$ with $\alpha=0$, $\beta=0$ and $\gamma=0$ only possesses an extra (local) conserved operator $\sigma_L^x$, since $-\sigma_L^z$ is absent, and the remaining five BCs stem from representations of either the free-ends or the periodic Temperley-Lieb algebra~\cite{tla,baxterbook,qpotts1,levy,saleur1}.  In other words, they represent typical BCs, in the sense that they may be identified as fixed points along boundary renormalization group (RG) flows, which are characterized in terms of the Affleck-Ludwig $g$ theorem that the boundary entropy is non-increasing along a boundary RG flow~\cite{ludwig}.

In addition,  the Hamiltonian (\ref{Ising}) is peculiar under one of four types of toroidal BCs, in the sense that it commutes with a variant of the translation operator $T_\mu$ ($\mu = p, ap, d, ad$). In fact, the Hamiltonian $H_{\rm}(\alpha,\beta,\gamma)$ commutes with $T_p$ under periodic BCs at $\alpha=1$, $\beta=0$ and $\gamma=1$, and it commutes with $T_{ap}=T_p \; \sigma_L^z$ under anti-periodic BCs at $\alpha=-1$, $\beta=0$ and $\gamma=1$. Here $T_P $ takes the form
\begin{equation}
T_p=\prod_{j=1}^{L-1}P_{jj+1},
\end{equation}
with
\begin{equation}
P_{jj+1}=\frac{1}{2}(\sigma_j^x\sigma_{j+1}^x+\sigma_j^y\sigma_{j+1}^y+\sigma_j^z\sigma_{j+1}^z+1). 
\end{equation}
Meanwhile, the Hamiltonian $H_{\rm}(\alpha,\beta,\gamma)$ commutes with $T_d=\nu_LT_p~d_L$ under duality-twisted BCs at $\alpha=0$,  $\beta=1$ and $\gamma=0$, and it commutes with $T_{ad}=\nu_LT_{ap}d_L$ under anti-duality-twisted BCs at $\alpha=0$, $\beta=-1$ and $\gamma=0$.  Here $\nu_L$ is a phase factor, defined as
\begin{equation*}
\nu_L=\exp(i(L+1)\pi/(4L-2)),
\end{equation*}
and
$d_L$ is defined as $d_L=g_{2L-2}\;g_{2L-1}$. In general, we introduce $d_j=g_{2j-2}\;g_{2j-1}$, with 
\begin{align}
g_{2j-2} &= - \frac {1+i}{2} + \frac {1-i}{2} \sigma_{j-1}^x \;\sigma_j^x,  \nonumber \\
g_{2j-1} &= - \frac {1+i}{2} + \frac {1-i}{2} \sigma_j^z. \label{gg}
\end{align}
A caveat is that $d_1$ involves $\sigma^x_0$ when $j=1$, which should be understood as $\sigma^x_L$.

In other words, the translation invariance is lacking for the Hamiltonian (\ref{Ising}) when $\alpha$, $\beta$ and $\gamma$ are away from the above four points.
In particular, the TFIM with free ends is not translation-invariant for each of the two choices of the boundary term $B_{L,1}(\alpha,\beta,\gamma)$ when $\alpha=0$, $\beta=0$ and $\gamma=1$ and $\alpha=0$, $\beta=0$ and $\gamma=0$. However, as we shall show below, this is so {\it only} when we restrict to the original Hilbert space.

For later uses, we introduce the KW unitary transformation $U_{KW}$, which takes the form
\begin{equation}
	U_{KW}=\prod_{j=1}^{L-1}\left[\left(\frac{1+i\sigma_j^z}{\sqrt{2}}\right)\left(\frac{1+i\sigma^x_j\sigma^x_{j+1}}{\sqrt{2}}\right)\right]\left(\frac{1+i\sigma^z_L}{\sqrt{2}}\right). 
	\label{KWS}
\end{equation}
Note that the squared KW unitary transformation $U_{KW}^2$ may be decomposed as follows
\begin{equation}
	U_{KW}^2= e^{i\pi L/2} \left(  T_p \frac{I+\eta}{2} +  T_{ap} \frac{I-\eta}{2} \right).
	\label{KWS}
\end{equation}
Here the Hilbert space  $\mathbb{C}^2_1 \otimes \mathbb{C}^2_2 \otimes \ldots \otimes \mathbb{C}^2_L$ itself is decomposed into the two sectors so that one is even and the other is odd under the ${\rm Z}_2$ symmetry operator $\eta$.

{\it The augmented Hilbert space and the augmented Hamiltonian.~--}  Note that $T_\mu$ for fixed $\mu$ does not commute with the Hamiltonian $H(\alpha,\beta,\gamma)$ for arbitrary $\alpha$, $\beta$ and $\gamma$. The repeated action of $T_\mu$ on the Hamiltonian $H(\alpha,\beta,\gamma)$ yield a sequence of the Hamiltonians $H_\mu^{(r)}(\alpha,\beta,\gamma)$, defined as 
\begin{equation*}
H_\mu^{(r)}(\alpha,\beta,\gamma) = T_\mu^r H_{\rm}(\alpha,\beta,\gamma) T_\mu^{-r}, 
\end{equation*}
with  $H_\mu^{(0)}(\alpha,\beta,\gamma) = H(\alpha,\beta,\gamma)$, where $r$ is a non-negative integer. Here we are concerned with
a specific feature of $T_\mu$: for a given $T_\mu$, there is a minimum integer $r_{\rm min}$ such that $T^{r_{\rm min}}_\mu$ commutes with  $H_{\rm}(\alpha,\beta,\gamma)$ for arbitrary $\alpha$, $\beta$ and $\gamma$. If so, we only need to restrict to $r=0,1,2,\ldots,r_{\rm min}-1\;{\rm mod}\;(r_{\rm min})$. It is readily seen that  both $T_p$ and $T_{ap}$ satisfy this constraint when $r_{\rm min}=L$, namely $T^L_p$ and $T^L_{ap}$ commute with  $H_{\rm}(\alpha,\beta,\gamma)$ for any $\alpha$, $\beta$ and $\gamma$, since  $T_p^L=I$ and $T_{ap}^L=\eta$.
In contrast, $T_d$ and $T_{ad}$  satisfy this constraint 
when $r_{\rm min}= 2(2L-1)$, namely $T_d^{2(2L-1)}$ and $T_{ad}^{2(2L-1)}$ commute with the Hamiltonian $H(\alpha,\beta,\gamma)$ for any $\alpha$, $\beta$ and $\gamma$,
since $T_d^{2(2L-1)}=\eta$ and $T_{ad}^{2(2L-1)}=-\eta$. 
Physically, this stems from the fact that, for arbitrary  $\alpha$, $\beta$ and $\gamma$,  the Hamiltonian $H_{\rm}(\alpha,\beta,\gamma)$ does not possess any extra nontrivial conserved quantity, except for the trivial identity operator $I$ and the ${\rm Z}_2$ symmetry operator $\eta$. 
However,  $T^L_p$ or $T^L_{ap}$ commutes with  $H(\alpha,\beta,\gamma)$ at $\alpha=1$, $\beta=0$ and $\gamma=1$ or $\alpha=-1$, $\beta=0$ and $\gamma=1$, respectively. Meanwhile,
$T^L_d$ or $T^L_{ad}$ commutes with  $H(\alpha,\beta,\gamma)$ at $\alpha=0$, $\beta=1$ and $\gamma=0$ or $\alpha=0$, $\beta=-1$ and $\gamma=0$, respectively.
Mathematically, we have $T_d^L=\exp(-i\pi/8)\xi_d$ and $T_{ad}^L=\exp(-i\pi/8)\xi_{ad}$, where 
\begin{equation*}
\xi_d =\exp \left(i\pi/8\right) \nu^L  ~d_1d_2 \ldots d_L,
\end{equation*}
and 
\begin{equation*}	 
\xi_{ad} = \exp \left(i\pi/8\right) \nu^L  d_1\; \sigma^z_2 d_2 \ldots \sigma^z_L d_Lf_L.
\end{equation*}
This follows from the fact that both duality-twisted and anti-duality-twisted BCs are integrable in the Yang-Baxter sense, because they arise from the periodic Temperley-Lieb algebra~\cite{levy}. Hence it is not surprising to see the presence of an extra nontrivial conserved quantity $\xi_d$ or $\xi_{ad}$ under either duality-twisted or anti-duality-twisted BCs.

In this construction, each of  $H_\mu^{(r)}(\alpha,\beta,\gamma)$ for fixed $\mu$ ($\mu=p$, $ap$, $\mu=d$ and $ad$) is on the {\it same} footing, as far as the thermodynamic limit is concerned.  As a result, one may introduce extra degrees of freedom, living in the auxiliary $r_{\rm min}$-dimensional complex vector space $\mathbb{C}^{r_{\rm min}}_A$, into the original Hilbert space $\mathbb{C}^2_1 \otimes \mathbb{C}^2_2 \otimes \ldots \otimes \mathbb{C}^2_L$ such that the augmented Hilbert space becomes $\mathbb{C}^2_1 \otimes \mathbb{C}^2_2 \otimes \ldots \otimes \mathbb{C}^2_L \otimes \mathbb{C}^{r_{\rm min}}_A$, where the subscript $A$ indicates that it is an auxiliary vector space. Hence we are capable of turning a non-translation-invariant Hamiltonian in the original Hilbert space into a translation-invariant Hamiltonian in the augmented Hilbert space.  Indeed, the augmented Hamiltonian $\mathscr{H}_\mu(\alpha,\beta,\gamma)$ is defined as
\begin{equation}
\mathscr{H}_\mu (\alpha,\beta,\gamma)= \sum_{r=0}^{r_{\rm min}-1} H_\mu^{(r)}(\alpha,\beta,\gamma) \vert r \rangle_A {}_A \langle r \vert, 
\end{equation}
where $\vert r \rangle_A$ ($r=0,1,2,\ldots,r_{\rm min}-1$) denote a set of the orthonormal basis states in the auxiliary vector space $\mathbb{C}^{r_{\rm min}}_A$. 
Similarly, we define  $\mathscr{T}_\mu \equiv \sum_{r=0}^{r_{\rm min}-1} T_\mu \vert r \rangle_A {}_A \langle r \vert$.
Now it is readily seen that 
\begin{equation}
	\mathscr{T}_\mu \mathscr{H}_\mu(\alpha,\beta,\gamma)\mathscr{T}_\mu^{-1} = \mathscr{C} \mathscr{H}_\mu(\alpha,\beta,\gamma) \mathscr{C}^{-1}, 
	\label{tc}
\end{equation} 
where $\mathscr{C}$ is a matrix in the auxiliary vector space $\mathbb{C}^{r_{\rm min}}_A$, defined as 
\begin{equation*}
\mathscr{C} \equiv \sum_{r=0}^{r_{\rm min}-2} \vert r+1 \rangle_A {}_A\langle r \vert + \vert 0 \rangle_A {}_A \langle r_{\rm min}-1 \vert. 
\end{equation*} 
Note that $\mathscr{C}$ commutes with $\mathscr{T}_\mu$. In other words, the augmented Hamiltonian $\mathscr{H}_\mu(\alpha,\beta,\gamma)$ commutes with a translation operator $\mathscr{T}_{\mu,o} = \mathscr{C}^{-1}\mathscr{T}_\mu$ in the augmented Hilbert space for arbitrary $\alpha$, $\beta$ and $\gamma$, if and only if $T^{r_{\rm min}}_\mu$ yields a unitary symmetry operator in the original Hilbert space. 

Mathematically, $\mathscr{C}$ generates a representation of the cyclic group ${\rm Z}_{r_{\rm min}}$ in the auxiliary vector space 
$\mathbb{C}^{r_{\rm min}}_A$, namely $\mathscr{C}^{r_{\rm min}} =I_A$, where $I_A$ denotes the identity operator in the augmented Hilbert space. Consequently, it {\it only} makes sense to speak of translation invariance  after a Hilbert space is specified.
This trick was introduced to find a translation-invariant matrix product state representation~\cite{exactmps}  for degenerate ground states arising from spontaneous symmetry breaking with type-B Goldstone modes~\cite{watanabe,NG}.
An important conclusion one may draw from this trick is that one specific Hamiltonian with the boundary term alone is not sufficient to characterize the TFIM at the self-dual point under various BCs. Instead, a sequence of the Hamiltonians $H^{(r)}_\mu(\alpha,\beta,\gamma)$ ($r=0,1,\ldots,r_{\rm min}-1$) are necessary, subject to the constraint that $T_\mu^{r_{\rm min}}$ commutes with $H(\alpha,\beta,\gamma)$ for arbitrary $\alpha$, $\beta$ and $\gamma$.

We emphasize that the presence of a translation operator $T_\mu$, subject to this constraint, is crucial for the construction of a lattice version of fusion rules under various BCs, with the free ends  as a representative case. 

{\it Lattice versions of fusion rules: a generic case.~--}  To proceed, we need to lift the identity operator $I$ and the ${\rm Z}_2$ symmetry operator $\eta$ in the original Hilbert space to the counterparts in the augmented Hilbert space, denoted as $I_A \equiv \sum_{r=0}^{L-1} I \; \vert r \rangle_A {}_A\langle r \vert$ and $\eta_A \equiv \sum_{r=0}^{L-1} \eta \; \vert r \rangle_A {}_A \langle r \vert$. Meanwhile, the unitary KW transformation $\mathscr{U}_{KW}$ is defined in the augmented Hilbert space, which is lifted from $U_{KW}$ defined in the original Hilbert space. Mathematically, we have $\mathscr{U}_{KW} \equiv \sum_r U_{KW}\; \vert r \rangle_A {}_A \langle r \vert$.

(i) For $H_p^{(r)}(\alpha,\beta,\gamma)$, we define the non-invertible KW duality symmetry 
\begin{equation*}
\mathscr{D}_{p,A} = \exp{(-i L \pi/4)}\;\sqrt {\mathscr{C}^{-1}  \mathscr{U}^2_{KW}}\;(I_A+\eta_A)/2. 
\end{equation*} 
Here $\sqrt {\mathscr{C}^{-1}  \mathscr{U}^2_{KW}}$ is the square root of $\mathscr{C}^{-1}  \mathscr{U}^2_{KW}$, which exists since $\mathscr{C}^{-1}  \mathscr{U}^2_{KW}$ is unitary.
It follows that the identity operator $I_A$, the ${\rm Z}_2$ symmetry operator $\eta_A$ and the non-invertible KW duality symmetry $\mathscr{D}_{p,A}$ yield a lattice analogue of fusion rules
\begin{align}
	\eta^2_A&=I_A,\nonumber\\
    \mathscr{T}_{p,o}^L &=I_A, \nonumber\\
	\eta_A \; \mathscr{T}_{p,o}&=\mathscr{T}_{p,o}\;\eta_A, \nonumber\\
	\eta_A \; \mathscr{D}_{p,A}&=\mathscr{D}_{p,A} \; \eta_A, \nonumber\\
	\mathscr{D}_{p,A}\; \mathscr{T}_{p,o}&=\mathscr{T}_{p,o} \; \mathscr{D}_{p,A}, \nonumber\\
	\mathscr{D}_{p,A}^2&=\frac{I_A+\eta_A}{2} \mathscr{T}_{p,o}. \label{fusion-generici}
\end{align}

(ii) For $H_{ap}^{(r)}(\alpha,\beta,\gamma)$, we define the non-invertible KW duality symmetry 
\begin{equation*} 
\mathscr{D}_{ap,A} = \exp{(-i L \pi/4)}\;\sqrt {\mathscr{C}^{-1} \mathscr{U}^2_{KW}}\;(I_A-\eta_A)/2.
\end{equation*} 
It follows that the identity operator $I_A$, the ${\rm Z}_2$ symmetry operator $\eta_A$ and the non-invertible KW duality symmetry $\mathscr{D}_{ap,A}$ yield a lattice analogue of fusion rules
\begin{align}
	\eta^2_A&=I_A,\nonumber\\
	\mathscr{T}_{ap,o}^L &=\eta_A, \nonumber\\
	\eta_A \; \mathscr{T}_{ap,o}&=\mathscr{T}_{ap,o}\;\eta_A, \nonumber\\
	\eta_A \; \mathscr{D}_{ap,A}&=\mathscr{D}_{ap,A} \; \eta_A, \nonumber\\
	\mathscr{D}_{ap,A}\; \mathscr{T}_{ap,o}&=\mathscr{T}_{ap,o} \; \mathscr{D}_{ap,A}, \nonumber\\
	\mathscr{D}_{ap,A}^2&=\frac{I_A-\eta_A}{2} \mathscr{T}_{ap,o}. \label{fusion-genericii}
\end{align}

(iii) For $H_d^{(r)}(\alpha,\beta,\gamma)$, we define the non-invertible KW duality symmetry 
\begin{align*}
\mathscr{D}_{d,A} =& \exp(-i\pi L/4) \sqrt {\nu_L}\;\sqrt {\mathscr{C}^{-1} \;\mathcal{d}_1\;\mathscr{U}^2_{KW}}\;(I_A+\eta_A)/2,
\end{align*}
where $\sqrt {\mathscr{C}^{-1} \mathcal{d}_1\mathscr{U}^2_{KW}}$ is the square root of $\mathscr{C}^{-1} \mathcal{d}_1 \mathscr{U}^2_{KW}$, which exists since $\mathscr{C}^{-1} \mathcal{d}_1 \mathscr{U}^2_{KW}$ is unitary. Here $\mathcal{d}_1$ is defined as  $\mathcal{d}_1 = \sum _{r} d_1\; \vert r \rangle_A {}_A \langle r \vert$. 
It follows that the identity operator $I_A$, the ${\rm Z}_2$ symmetry operator $\eta_A$ and the non-invertible KW duality symmetry $\mathscr{D}_{p,A}$ yield a lattice analogue of fusion rules
\begin{align}
	\eta^2_A&=I_A,\nonumber\\
	\mathscr{T}_{d,o}^{2(2L-1)} &=\eta_A, \nonumber\\
	\eta_A \; \mathscr{T}_{d,o}&=\mathscr{T}_{d,o}\;\eta_A, \nonumber\\
	\eta_A \; \mathscr{D}_{d,A}&=\mathscr{D}_{d,A} \; \eta_A, \nonumber\\
	\mathscr{D}_{d,A}\; \mathscr{T}_{d,o}&=\mathscr{T}_{d,o} \; \mathscr{D}_{d,A}, \nonumber\\
	\mathscr{D}_{d,A}^2&=\frac{I_A+\eta_A}{2} \mathscr{T}_{d,o}. \label{fusion-genericiii}
\end{align}

(iv) For $H_{ad}^{(r)}(\alpha,\beta,\gamma)$, we define the non-invertible KW duality symmetry 
\begin{align*}
\mathscr{D}_{ad,A} =  \exp(-i\pi L/4) \sqrt {\nu_L}\;\sqrt { \mathscr{C}^{-1}\;\mathcal{d}_{1,ad}\;\mathscr{U}^2_{KW}}\;(I_A-\eta_A)/2.
\end{align*}
Here $\mathcal{d}_{1,ad}$ is defined as  $\mathcal{d}_{1,ad} = \sum _{r} d_{1,ad}\; \vert r \rangle_A {}_A \langle r \vert$, where $d_{1,ad}$
takes the same form as $d_{1}$ in Eq.~(\ref{gg}) when $j=1$,  with the difference that $\sigma^x_0$ should be understood as $-\sigma^x_L$. Note that $\sqrt {\mathscr{C}^{-1}\mathcal{d}_{1,ad}\mathscr{U}^2_{KW}}$ is the square root of $\mathscr{C}^{-1}\mathcal{d}_{1,ad}\mathscr{U}^2_{KW}$, which exists since $\mathscr{C}^{-1}\mathcal{d}_{1,ad}\mathscr{U}^2_{KW}$ is unitary.
It follows that the identity operator $I_A$, the ${\rm Z}_2$ symmetry operator $\eta_A$ and the non-invertible KW duality symmetry $\mathscr{D}_{ad,A}$ yield a lattice analogue of fusion rules
\begin{align}
	\eta^2_A&=I_A,\nonumber\\
	\mathscr{T}_{ad,o}^{2(2L-1)} &=-\eta_A, \nonumber\\
	\eta_A \; \mathscr{T}_{ad,o}&=\mathscr{T}_{ad,o}\;\eta_A, \nonumber\\
	\eta_A \; \mathscr{D}_{ad,A}&=\mathscr{D}_{ad,A} \; \eta_A, \nonumber\\
	\mathscr{D}_{ad,A}\; \mathscr{T}_{ad,o}&=\mathscr{T}_{ad,o} \; \mathscr{D}_{ad,A}, \nonumber\\
	\mathscr{D}_{ad,A}^2&=\frac{I_A-\eta_A}{2} \mathscr{T}_{ad,o}. \label{fusion-genericiv}
\end{align}

As already stressed, it works for any BCs, as long as the boundary term is chosen to ensure that the (on-site) ${\rm Z}_2$ symmetry operator $\eta$ is retained for arbitrary $\alpha$, $\beta$ and $\gamma$. Hence our construction resolves an apparent contradiction with the basic physical requirement that  all possible BCs should yield a converging result in the thermodynamic limit. In addition, it also clarifies the essential difference between topological and non-topological defects.  Note that the location of a topological defect is arbitrary and can be varied by conjugating the defect Hamiltonian with local unitary operators in the original Hilbert space, in contrast to non-topological defects. Indeed, it is always possible to construct translation symmetry operators in the original Hilbert space for topological defects, whereas translation symmetry operators only exist in the augmented Hilbert space for non-topological defects.

{\it Lattice versions of fusion rules: toroidal BCs.~--} As already mentioned above, a variant of the translation symmetry operator emerges for the TFIM under each of four types of toroidal BCs, which appear to be periodic and anti-periodic BCs for $\alpha=\pm 1$, $\beta=0$ and $\gamma=1$ and duality-twisted and anti-duality-twisted Bcs for $\alpha=0$, $\beta=\pm 1$ and $\gamma=0$.  Given that the above construction applies to any  $\alpha$, $\beta$ and $\gamma$, one may wonder what happens to the TFIM under four types of toroidal BCs. From Eq.~(\ref{tc}), it is readily seen that $\mathscr{C}$ becomes conserved for the augmented Hamiltonian defined in the augmented Hilbert space if and only if $T_\mu$ commutes with the Hamiltonian  defined in the original Hilbert space.  This is exactly what we desire for the TFIM model under four types of toroidal BCs, given that $T_p$ and $T_{ap}$ commute with the Hamiltonian under periodic and anti-periodic BCs, and $T_d$ and $T_{ad}$ commute with the Hamiltonian under duality-twisted and anti-duality-twisted BCs, respectively. Hence one may remove $\mathscr{C}$ from $\mathscr{D}_{\mu,A}$ ($\mu=p, ap, d, ad$), when we restrict to four types of toroidal BCs. As a result, the auxiliary vector space $C^{r_{\rm min}}_A$ is manifested in such a way that the same lattice version of fusion rules is replicated $r_{\rm min}$ times for each of four types of toroidal BCs.  Consequently, we are able to get rid of this repetition and pull back the lattice versions of fusion rules in the augmented Hilbert space to those in the original Hilbert space. We are thus led to the lattice versions of 
fusion rules for four types of toroidal BCs.

(i) For the TFIM under periodic BCs, one may define
the non-invertible KW duality symmetry $D_p$
\begin{equation}
D_p =\exp{(-i L \pi/4)}\; U_{KW} \; \frac{I+\eta}{2}. \nonumber \\
\label{syising}
\end{equation}
As a result, the identity operator $I$,  the ${\rm Z}_2$ symmetry operator $\eta$ and  the non-invertible KW duality symmetry $D_p$ satisfy
\begin{align}
\eta^2&=I,\nonumber\\
T_p^L&=I,\nonumber\\
\eta \;T_p&=T_p \; \eta, \nonumber\\
\eta\;  D_p&=D_p \; \eta, \nonumber\\
D_p \; T_p&=T_p \; D_p, \nonumber\\
D_p^2&=\frac{I+\eta}{2} \; T_p. \label{fusioni}
\end{align}

(ii) For the TFIM under  anti-periodic BCs, one may define
the non-invertible KW duality symmetry $D_{ap}$
\begin{equation}
 D_{ap}=\exp{(-i L\pi /4)}\;U_{KW}\;\frac{I-\eta}{2}. \nonumber \\
 \label{syising}
\end{equation}
As a result, the identity operator $I$,  the ${\rm Z}_2$ symmetry operator $\eta$ and  the non-invertible KW duality symmetry $D_{ap}$ satisfy
\begin{align}
\eta^2&=I,\nonumber\\
T_{ap}^L&=\eta,\nonumber\\
\eta \; T_{ap}&=T_{ap} \; \eta, \nonumber\\
\eta \;  D_{ap}&=D_{ap} \; \eta, \nonumber\\
D_{ap} \; T_{ap}&=T_{ap} \; D_{ap}, \nonumber\\
D_{ap}^2&=\frac{I-\eta}{2}\; T_{ap}. \label{fusionii}
\end{align}

(iii) For the TFIM under duality-twisted BCs, one may define
the non-invertible KW duality symmetry $D_d$
\begin{equation}
	D_d = \exp(-i\pi L/4) \sqrt {\nu_L}\; \sqrt {d_1 U^2_{KW}} \; \frac{I+\eta}{2}, \nonumber \\
	\label{syising}
\end{equation}
where $\sqrt {d_1 U^2_{KW}}$ is the square root of $d_1 U^2_{KW}$, which exists since $d_1 U^2_{KW}$ is unitary.
Hence the identity operator $I$,  the ${\rm Z}_2$ symmetry operator $\eta$ and  the non-invertible KW duality symmetry $D_d$ satisfy
\begin{align}
	\eta^2&=I,\nonumber\\
    T_d^{2(2L-1)}&=\eta, \nonumber\\
	\eta \;T_d&=T_d \; \eta, \nonumber\\
	\eta\;  D_d&=D_d \; \eta, \nonumber\\
	D_d \; T_d&=T_d \; D_d, \nonumber\\
	D_d^2&=\frac{I+\eta}{2} \; T_d. \label{fusioniii}
\end{align}
However, the presence of an extra symmetry operator $\xi_d$ requires to enlarge this operator algebra. We have
$\xi^2_d= \frac{\sqrt {2}}{2} (1+i~\eta)T_d$, where $\xi_d$ commutes with $\eta$, $T_d$ and $D_d$, namely
 $\xi_d\eta=\eta\xi_d$, $\xi_d T_d=T_d\xi_d$ and $\xi_d D_d=D_d\xi_d$,
in addition to the known relation $T_d^L=\exp(-i\pi/8)~\xi_d$. Note that  $T_d^{2(2L-1)}=(1-i)\;T_d^{2L-1}+i$.

(iv) For the TFIM under anti-duality-twisted BCs, one may define
the non-invertible KW duality symmetry $D_{ad}$
\begin{equation}
	D_{ad} = \exp(-i\pi L/4) \sqrt {\nu_L}\;\sqrt {d_{1,ad} U^2_{KW}} \; \frac{I-\eta}{2}, \nonumber \\
	\label{syising}
\end{equation}
where $\sqrt {d_{1,ad} U^2_{KW}}$ is the square root of $d_{1,ad} U^2_{KW}$, which exists since $d_{1,ad} U^2_{KW}$ is unitary.
Hence the identity operator $I$,  the $Z_2$ symmetry operator $\eta$ and  the non-invertible KW duality symmetry $D_{ad}$ satisfy
\begin{align}
	\eta^2&=I,\nonumber\\
	T_{ad}^{2(2L-1)}&=-\eta,\nonumber\\
	\eta \;T_{ad}&=T_{ad} \; \eta, \nonumber\\
	\eta\;  D_{ad}&=D_{ad} \; \eta, \nonumber\\
	D_{ad} \; T_{ad}&=T_{ad} \; D_{ad}, \nonumber\\
	D_{ad}^2&=\frac{I-\eta}{2} \; T_{ad}. \label{fusioniv}
\end{align}
However, the presence of an extra symmetry operator $\xi_{ad}$ requires to enlarge this operator algebra. We have 
$\xi^2_{ad}= \frac{\sqrt {2}}{2} (1-i~\eta)T_{ad}$, where $\xi_{ad}$ commutes with $\eta$, $T_{ad}$ and $D_{ad}$, namely
$\xi_{ad}\eta=\eta\xi_{ad}$, $\xi_{ad} T_{ad}=T_{ad}\xi_{ad}$ and $\xi_{ad} D_{ad}=D_{ad}\xi_{ad}$, in addition to the known relation
$T_{ad}^L= \exp(-i\pi/8)~\xi_{ad}$.
Note that $T_{ad}^{2(2L-1)}=(1-i)\;T_{ad}^{2L-1}+i$.

Consequently, these four  types of toroidal BCs are peculiar, in the sense that, in each case,  the Hamiltonian is translation-invariant as a result of the compensation arising from conjugating the defect Hamiltonian with local unitary operators
in the original Hilbert space. In other words, they are qualified as topological defects.  
Here we stress that for $T_d$ and $T_{ad}$, extra care must be taken when $r_{\rm min}$ is determined, given that $r_{\rm min}$ is defined to be the minimum integer such that $T^{r_{\rm min}}_d$ commutes with  $H_{\rm}(\alpha,\beta,\gamma)$ for arbitrary $\alpha$, $\beta$ and $\gamma$. Since  $H_{\rm}(\alpha,\beta,\gamma)$ for arbitrary $\alpha$, $\beta$ and $\gamma$ is not integrable, it is impossible to possess any other extra symmetry operator other than the trivial identity operator and the $Z_2$ symmetry operator $\eta$. This is in sharp contrast to the TFIM model under duality-twisted or anti-duality-twisted BCs,  given that each of the two BCs is integrable, so an extra (unitary) symmetry operator $\xi_d$ or $\xi_{ad}$ emerges. Hence the requirement that $T^{r_{\rm min}}_d$ commutes with  $H_{\rm}(\alpha,\beta,\gamma)$ for arbitrary $\alpha$, $\beta$ and $\gamma$ is equivalent to stating that $T^{r_{\rm min}}_d$ is essentially $\eta$ itself.  As a result, a variant of the translation symmetry operator is involved in a lattice analogue of fusion rules in the original Hilbert space. We are thus capable of reproducing the results, constructed by Seiberg, Seifnashri and Shao~\cite{seiberg}, for the TFIM under periodic and anti-periodic BCs. However, our construction reveals that non-invertible KW duality symmetries also exist for the TFIM under duality-twisted and anti-duality-twisted BCs,  in contrast to their claim that the KW duality symmetry is unitary, so it is invertible for duality-twisted BCs~\cite{seiberg}. In fact, they have misinterpreted the counterpart of the extra (unitary) symmetry operator $\xi_d$ as the invertible KW duality symmetry. Their symmetry operator algebra is thus incomplete (a brief discussion for a unitary equivalence between their Hamiltonian and ours is presented in the Supplementary Material).

{\it Summary and outlook.~--} 
We have re-visited lattice versions of fusion  rules arising from non-invertible KW duality symmetries for the TFIM at the self-dual point under toroidal BCs. 
The construction has been extended to non-toroidal BCs, as long as the resultant Hamiltonian commutes with the ${\rm Z}_2$ symmetry operator.  
As it turns out, there are always four lattice versions of fusion rules at a generic point in the parameter space $(\alpha,\beta, \gamma)$, namely (\ref{fusion-generici}), (\ref{fusion-genericii}), (\ref{fusion-genericiii}) and (\ref{fusion-genericiv}) in the augmented Hilbert space. In addition, one more lattice version in the original Hilbert space occurs at the four points $(\pm 1,0,1)$ and $(0,\pm 1,0)$, namely (\ref{fusioni}) and (\ref{fusionii}) for  periodic and anti-periodic BCs, and (\ref{fusioniii}) and (\ref{fusioniv}) for duality-twisted and anti-duality-twisted BCs.
In principle, our construction may be extended to the $q$-state QP model under various BCs, which is under active investigation. Here we note that the three-state QP model has been investigated under periodic BCs~\cite{sierra}.

In closing, we emphasize that a non-invertible KW duality symmetry is a feature for the Hamiltonian itself. 
The usefulness of this observation  lies in the fact that non-invertible KW duality symmetries will play a crucial role in understanding the physics underlying many well-studied quantum many-body lattice systems, with their ground state sub-manifolds being either conformally or non-conformally invariant. Here we remark that the AF or FM $q$-state QP model has been revealed to be unitarily equivalent to 
the FM or AF ${\rm SU}(n)$ spin-$s$ chain, where $q=n^2$ and $n=2s+1$, as long as the size of  the latter is doubled~\cite{ue1}, if one restricts to the free-ends BCs.
In particular, the FM or AF four-state QP model is unitarily equivalent to the FM or AF double TFIM, with the symmetry group ${\rm Z}_2 \times {\rm Z}_2$. The latter in turn is identified as the FM or AF Ashkin-Teller model at a particular point, which is located at one endpoint of the critical line in the FM case~\cite{ashkinteller,ashkinteller1}. Moreover, the FM or AF double TFIM is
unitarily equivalent to the AF or FM ${\rm SU}(2)$ spin-$1/2$ Heisenberg model. This implies that our construction may be extended to the AF or FM ${\rm SU}(2)$ spin-$1/2$ Heisenberg model~\cite{noninveriblessb}  in particular and many other Temperley-Lieb integrable models in general, which in turn lead to the projected Green parafermion states~\cite{zhougreen}.

{\it Acknowledgment.~--} We thank Ian P. McCulloch for bringing our attention to non-invertible symmetries. We also thank Murray T. Batchelor,  John O. Fj{\ae}restad and Ian P. McCulloch for enlightening discussions.

\newpage
\newpage
\section*{Supplementary Material}
\twocolumngrid
\setcounter{equation}{0}
\renewcommand{\theequation}{S\arabic{equation}}

The Hamiltonian of the TFIM under duality-twisted BCs (at the self-dual point), adopted by Seiberg, Seifnashri and Shao~\cite{seiberg}, takes the form
\begin{equation}
	H_g=-\sum_{j=2}^{L} \left (\sigma_{j-1}^z \sigma_{j}^z +  \sigma_{j}^x \right)-\sigma^z_L\; \sigma^x_1.  \label{Ising-ue} 
\end{equation}
Note that the Hamiltonian $H_g$ in Eq.~(\ref{Ising-ue} ) is unitarily equivalent to the Hamiltonian (\ref{Ising}) under duality-twisted BCs, with the boundary term  $B_{L,1}(\alpha,\beta,\gamma) =\alpha\sigma_L^x \sigma_{1}^x+\beta \sigma_L^y \sigma_{1}^x+\gamma\sigma_L^z$ at $\alpha=0$, $\beta=1$ and $\gamma=0$, which is denoted as $H_d$ below for brevity. Mathematically, we have $H_z = V  H_g V^\dagger$, where $V$ is the unitary matrix. A detailed calculation shows that $V$ takes the form 
\begin{equation}
V= K_1\;K_2, 
\end{equation}
where $K_1$ and $K_2$ are
\begin{align*}
	K_1 =& \frac {\sqrt 2}{4} \left( 1+\sigma_1^z+\sigma_L^z - \sigma_1^z\;\sigma_L^z \right) \left( \exp (i\;\frac {\pi}{4}+\exp (-i \frac {\pi}{4} ) \;\sigma_L^z \right),  \nonumber\\
	K_2 =& \exp (i \frac {\pi}{2} \sigma_L^x) \prod_{j=1}^L \exp (i\;\frac {\pi}{4}\;\sigma_j^y).
\end{align*}
Meanwhile, the ${\rm Z}_2$ symmetry operator $\eta$ is converted into the symmetry operator $\eta_D$ (up to a prefactor $-i$), where $\eta_D= \sigma^x_L \ldots \sigma^x_2\sigma^z_1 \sigma^x_1$~\cite{seiberg}. Namely, we have  $\eta_D = -i  V^\dagger  \eta V$. As argued in Ref.~\cite{ue1}, unitarily equivalent models in the same family are essentially identical, but often appear in different guises. However, the situation here is quite simple, since both $\eta$ and $\eta_D$ are on-site operators. In fact, the lattice version of fusion rules (\ref{fusioniv}) is retained, if one exploits the unitary transformation $V$ and its Hermitian conjugation $V^\dagger$ to map the identity operator $I$,  the ${\rm Z}_2$ symmetry operator $\eta$ and the non-invertible KW duality symmetry $D_d$ for the Hamiltonian $H_z$ to their counterparts for the Hamiltonian $H_g$, in addition to $T_d$. Actually,  this mapping for $\eta$ has been written out above, and it is trivial for the identity operator $I$. 
Define $T_D^{-1}=\exp(i\pi/(8L-4))V^\dagger T_d  V$, where $T_D$ has been introduced in Ref.~\cite{seiberg}. As a result, $T_d^{2(2L-1)}$ becomes 
$T_D^{-2(2L-1)}$. 
However, Seiberg, Seifnashri and Shao~\cite{seiberg} misinterpreted a counterpart of the extra (unitary) symmetry operator $\xi_d$ as the invertible KW duality symmetry. Hence the symmetry operator algebra they have constructed is incomplete for the TFIM under duality-twisted BCs. In other words, they did not introduce the counterpart for the non-invertible KW duality symmetry $D_d$.

Here we simply define the mapping for the extra (unitary) symmetry operator $\xi_D$ and the non-invertible KW duality symmetry $D_d$ to be  $\xi_D=  \exp(iL\pi/(8L-4)) V^\dagger \xi_d V$
and $D_D = \exp(i\pi/(16L-8)) V^\dagger D_d V$. Hence the operator algebra (\ref{fusioniii}) is mapped to the following form
\begin{align}
	\eta^2_D&= -I,\nonumber\\
	T_D^{-2(2L-1)}&=-\eta_D, \nonumber\\
	\eta_D \;T_D&=T_D \; \eta_D, \nonumber\\
	\eta_D\;  D_D&=D_D \; \eta_D, \nonumber\\
	D_D \; T_D&=T_D \; D_D, \nonumber\\
	D_D^2&=\frac{I+i\eta_D}{2} \; T_D^{-1}. \label{fusioniii-map}
\end{align}
Note that the presence of the extra symmetry operator $\xi_D$ requires to enlarge the above operator algebra.  We have $T_D^{-L}=\exp(-i\pi/8)~\xi_D$. It is readily seen that
$\xi^2_D= \frac{1+i}{2} (1-\eta_D)T_D^{-1}$, where $\xi_D$ commutes with $\eta_D$, $T_D$ and $D_D$.
In particular, the relation  
\begin{equation*}
T_d^{2(2L-1)}=(1-i)\;T_d^{2L-1}+i
\end{equation*}
becomes
\begin{equation*}
T_D^{2(2L-1)}=\sqrt{2} \;T_D^{2L-1}-1.  
\end{equation*}
In other words, we have reproduced the above relation for $T_D$, which has been derived in Ref.~\cite{fendley}.

An important lesson one may learn from this example for a unitarily equivalent family is that different members are subject to different symmetry operators that are connected by the unitary transformation $V$ and its Hermitian conjugation $V^\dagger$, as exemplified by $\eta$ and $\eta_D$. In fact, many examples show that a unitary transformation $V$ connecting two members in a generic unitarily equivalent family is highly non-local such that an on-site symmetry generator is mapped to a non-on-site symmetry generator~\cite{ue1}.

On the other hand, it is important to keep the {\it same} symmetry operator when different boundary terms are introduced to the same bulk Hamiltonian, when a lattice version of fusion rules as a symmetry operator algebra is discussed. In principle, one may introduce a boundary term that breaks the symmetry group in the 
bulk Hamiltonian, even if this boundary term itself is integrable, in the sense that it satisfies a modified form of the TL algebra~\cite{levy}. If so, it is still possible to construct a generalized symmetry operator algebra to include non-invertible KW duality symmetries.

\end{document}